\documentclass[preprint,showpacs,preprintnumbers,amsmath,amssymb,nofootinbib]{revtex4}
\usepackage{bbm}
\usepackage{amsfonts}
\usepackage{booktabs}
\usepackage{mathrsfs}
\usepackage{epsfig}
\usepackage{graphicx}
\usepackage{dcolumn}
\usepackage{bm}
\usepackage{amsmath}
\usepackage{slashed}

\let\jnfont=\rm
\def\NPB#1,{{\jnfont Nucl.\ Phys.\ B }{\bf #1},}
\def\PLB#1,{{\jnfont Phys.\ Lett.\ B }{\bf #1},}
\def\EPJC#1,{{\jnfont Eur.\ Phys.\ Jour.\ C }{\bf #1},}
\def\PRD#1,{{\jnfont Phys.\ Rev.\ D }{\bf #1},}
\def\PRL#1,{{\jnfont Phys.\ Rev.\ Lett.\ }{\bf #1},}
\def\MPLA#1,{{\jnfont Mod.\ Phys.\ Lett.\ A }{\bf #1},}
\def\JPG#1,{{\jnfont J.\ Phys.\ G}{\bf #1},}
\def\CTP#1,{{\jnfont Commun.\ Theor.\ Phys.\ }{\bf #1},}
\def\ZPC#1,{{\jnfont Z.\ Phys.\ C }{\bf #1},}
\def\JHEP#1,{{\jnfont JHEP \ }{\bf #1},}
\def\Rv{\not{\hbox{\kern-1pt $R$}}}
\def\p{\not{\hbox{\kern-3pt $p$}}}

\begin{document}
\preprint{\parbox{1.2in}{\noindent arXiv:}}

\title{Current experimental bounds on stop mass in natural SUSY}

\author{Chengcheng Han$^1$, Ken-ichi Hikasa$^{2}$, Lei Wu$^{3}$, Jin Min Yang$^1$, Yang Zhang$^{1,4}$
        \\~ \vspace*{-0.3cm} }
\affiliation{ $^1$ State Key Laboratory of Theoretical Physics, Institute of Theoretical Physics,
 Academia Sinica, Beijing 100190, China\\
$^2$ Department of Physics, Tohoku University, Sendai 980-8578, Japan\\
$^3$ ARC Centre of Excellence for Particle Physics at the Terascale, School of Physics,
The University of Sydney, NSW 2006, Australia\\
$^4$ Physics Department, Henan Normal University, Xinxiang 453007, China
 \vspace*{1.5cm}}

\begin{abstract}
Motivated by the recent progress of direct search for the productions of stop pair and sbottom
pair at the LHC, we examine the constraints of the search results on the stop ($\tilde{t}_1$) mass
in natural SUSY. We first scan the parameter space of natural SUSY in the framework of MSSM,
considering the constraints from the Higgs mass, B-physics and electroweak precision measurements.
Then in the allowed parameter space we perform a Monte Carlo simulation for stop pair production
followed by $\tilde{t}_{1} \to t \tilde{\chi}_{1}^{0}$ or $\tilde{t}_{1} \to b \tilde{\chi}_{1}^{+}$
and sbottom pair production followed by $\tilde{b}_{1} \to b \tilde{\chi}_{1}^{0}$ or
$\tilde{b}_{1} \to t \tilde{\chi}_{1}^{-}$.
Using the combined results of ATLAS with 20.1 fb$^{-1}$ from
the search of $\ell+{\rm jets}+\slashed E_{T}$, hadronic $t\bar{t}+\slashed E_{T}$ and $2b+\slashed E_{T}$,
we find that a stop lighter than 600 GeV can be excluded at 95\% CL
in this scenario.
\end{abstract}
\maketitle

\section{INTRODUCTION}
Observation of a SM-like Higgs boson around 125 GeV by the ATLAS and CMS
Collaborations \cite{ATLAS-CMS} has had a great impact on the
minimal supersymmetric standard model (MSSM), in which the naturalness
problem is a key issue \cite{mssm-ft,finetuning,nsusy1,nsusy2,nsusy3,nsusy4,nsusy5}.
For the electroweak naturalness the stops play an important role by cancelling the top quark loop,
and thus the LHC search for the stop pair production
will shed light on this question in the MSSM.
The ATLAS and CMS collaborations have also made searches for supersymmetric particles with about 20 fb$^{-1}$ of data \cite{exp-incl-ew,exp-st,exp-sb}.
The negative results of sparticle search may indicate the natural SUSY scenario \cite{nsusy1,nsusy2,nsusy3,nsusy4,nsusy5},
in which the first two generations of squarks are heavy and the third generation of squarks are light.

In the framework of MSSM, the fine-tuning can be estimated by the Barbieri-Giudice measure \cite{finetuning}
\begin{eqnarray}
&& \Delta=\max \{\Delta_{p_i}\} ,\nonumber \\
&& \Delta_{p_i}=\left|\frac{\partial\log M_Z^2}{\partial\log p_i^2}\right|\, .
\label{finetune}
\end{eqnarray}
where $p_i$ denotes the soft SUSY-breaking terms that enter the minimization conditions
of the Higgs potential \cite{mz}
\begin{eqnarray}
\frac{M^2_{Z}}{2}=\frac{(m^2_{H_d}+\Sigma_{d})-(m^2_{H_u}+
\Sigma_{u})\tan^{2}\beta}{\tan^{2}\beta-1}-\mu^{2},
\label{minimization}
\end{eqnarray}
with  $\tan\beta\equiv v_u/v_d$,
$m^2_{H_d}$ and $m^2_{H_u}$ representing the weak scale soft SUSY breaking masses of
the Higgs fields, $\mu$ being the higgsino mass parameter,
$\Sigma_{u}$ and $\Sigma_{d}$ being the radiative corrections to the tree-level Higgs potential.
In order to maintain naturalness, the parameters $\mu$, $m_{H_u}$ and $\Sigma_{u}$
are expected to be at weak scale, which will impose a constraint on the spectrum
of MSSM \cite{nsusy3}
\begin{eqnarray}
\label{eq:natural_spectrum}
&& \mu \lesssim 200\;{\rm GeV}\left(\frac{m_{h}}{125\;{\rm GeV}}\right)
\sqrt{\frac{\Delta}{5}} \; ,\\
&&\sqrt{ m_{\tilde{t}_1}^2+ m_{\tilde{t}_2}^2+A^{2}_{t}}\lesssim  600\;{\rm GeV} \sin\beta\;
 L_\Lambda \left(\frac{m_{h}}{125\;{\rm GeV}}\right)\sqrt{\frac{\Delta}{5}} \; , \\
&& M_{3} \lesssim  900\;{\rm GeV} \sin\beta \; L_\Lambda^2
\left(\frac{m_{h}}{125\;{\rm GeV}}\right)\sqrt{\frac{\Delta}{5}} \; ,
\end{eqnarray}
with $L_\Lambda = \sqrt{3/\log\left(\Lambda/{\rm TeV}\right)}$ coming from the
leading-log approximation of the RGE \cite{nsusy3}.
Of course, in a given model like the MSSM, CMSSM or natural SUSY, the fine-tuning extent
depends on the parameters appearing in the left sides of Eqs.~(3--5), which
can be checked by examining the parameter space allowed by current experiments
\cite{mssm-cmssm,wu,nsusy-stop,hanz,drees,met,bai,combine,boosted,nsusy-dm}.

In this work we focus on natural SUSY and examine the constraints of
the LHC search results of stop and sbottom \cite{exp-st,exp-sb}.
Note that some studies in this
aspect have been performed in the literature \cite{no-at,alphat}.
However, the study in \cite{no-at} assumed a special
scenario with $A_t=0$ and $\tan\beta=10$ and, as a result, the stop decay
is much simplified, while the analysis in \cite{alphat} did not consider
the ATLAS sbottom search results. In our study we will consider a more general natural
SUSY with more experimental constraints.
We will first scan the parameter space of natural SUSY in the framework of MSSM,
considering the constraints from the LHC Higgs mass, B-physics and electroweak
precision measurements, and then in the allowed parameter space we will perform
a detailed Monte Carlo simulation for stop pair production
with $\tilde{t}_{1} \to t \tilde{\chi}_{1}^{0}$ or $\tilde{t}_{1} \to b \tilde{\chi}_{1}^{+}$
and sbottom pair production with $\tilde{b}_{1} \to b \tilde{\chi}_{1}^{0}$ or
$\tilde{b}_{1} \to t \tilde{\chi}_{1}^{-}$ at the LHC, applying the same cuts
as in the ATLAS searches.
Finally, using the combined results of ATLAS with 20.1 fb$^{-1}$ from
the search of $\ell+{\rm jets}+\slashed E_{T}$, hadronic $t\bar{t}+\slashed E_{T}$ and $2b+\slashed E_{T}$,
we will figure out the allowed range for the stop in natural SUSY.

We organize the paper as follows. In Sec. II, we scan the parameter space and
present the spectrum of natural SUSY allowed by the Higgs mass, B-physics and
electroweak precision measurements.
In Sec. III we simulate the direct stop/sbottom pair productions and examine their
constraints on the parameter space of natural SUSY.
Finally, we draw our conclusion in Sec. IV.

\section{A scan of the parameter space}
In accordance with the definition of natural SUSY,
$\mu$ should have a value at weak scale and the third generation of squarks
should not be too far above TeV scale. So we scan these parameters in the
following ranges
\begin{eqnarray}
100~{\rm GeV} \le \mu \le 200~{\rm GeV},
~~~100~{\rm GeV} \le (m_{\tilde{t}_L}= m_{\tilde{b}_L}, ~m_{\tilde{t}_R})
\le 2~{\rm TeV}.
\end{eqnarray}
In natural SUSY the gluino mass is supposed to be below several TeV. Since in our study
we focus on stop/sbottom direct searches, the gluino mass is not a sensitive parameter,
which is fixed at 2 TeV in our scan.  The sleptons and
first two generations of squarks in natural SUSY are supposed to be heavy, which are all
fixed at 5 TeV. For the electroweak gaugino masses,
we use the grand unification relation $M_1:M_2 = 1:2$ and take $M_1 =1$ TeV (in natural SUSY
$M_1$ is usually supposed to be larger than $\mu$ and the lightest neutralino is
higgsino-like).
Further, we assume $A_{t}=A_{b}$ and $m_{\tilde{b}_R}=m_{\tilde{t}_R}$, which in our scan vary in
the ranges
\begin{eqnarray}
&& |A_{t}| \le 3~{\rm TeV}, ~~1 \le \tan\beta \le 60, ~~90~{\rm GeV} \le M_A \le 1~{\rm TeV},
\end{eqnarray}
where the lower bound of $M_A$ is indicated from LEP and B-decay constraints.
Since the component of the stop ($\tilde{t}_1$) will affect its decay modes,
we scan the parameter space in two scenarios: (I) $m_{\tilde{t}_R}$ is lighter than $m_{\tilde{t}_L}$;
(II) $m_{\tilde{t}_L}$ is lighter than $m_{\tilde{t}_R}$.

In our scan we consider the following constraints:
\begin{itemize}
\item[(1)] We require the SM-like Higgs mass in the range of 123--127 GeV.
We calculate the Higgs mass with \textsf{FeynHiggs2.8.9} \cite{feynhiggs} and impose the experimental
constraints from LEP, Tevatron and LHC with \textsf{HiggsBounds-3.8.0} \cite{higgsbounds}.
\item[(2)] We require natural SUSY to satisfy various B-physics bounds at 2$\sigma$ level.
We use the package of \textsf{SuperIso v3.3} \cite{superiso} to implement the constraints, including
$B\rightarrow X_s\gamma$ and the latest measurements of $B_s\rightarrow \mu^+\mu^-$,
$B_d\rightarrow X_s\mu^+\mu^-$ and $ B^+\rightarrow \tau^+\nu$.
\item[(3)] We require the SUSY prediction of the precision electroweak observable such as
$\rho_l$, $\sin^2 \theta_{\rm eff}^l$, $m_W$ and $R_b$ \cite{rb} to be within the $2\sigma$ ranges
of the experimental values.
\item[(4)] We require the thermal relic density of the lightest neutralino (as the dark matter candidate)
is below the 2$\sigma$ upper limit of the Planck value \cite{planck}.
We use the code \textsf{MicrOmega v2.4} \cite{micromega} to calculate the relic density.
\end{itemize}

\begin{figure}[h]
\centering
\includegraphics[width=5.5in,height=3in]{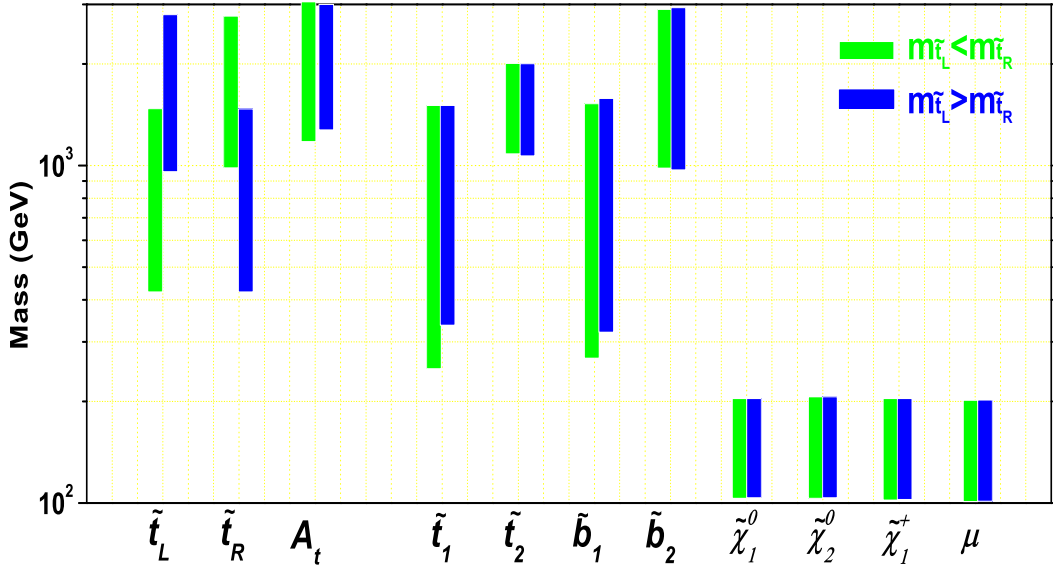}
\caption{Sparticle mass spectrum in the parameter space allowed by constraints (1--4) in natural SUSY.}
\label{spectrum}
\end{figure}
In Fig.~\ref{spectrum} we show the sparticle spectrum in the parameter space allowed by constraints (1-4)
in natural SUSY. From the figure we can find the following features and implications.
Firstly, for both $m_{\tilde t_L}<m_{\tilde t_R}$ and  $m_{\tilde t_L}>m_{\tilde t_R}$,
the stop ($\tilde{t}_1$) mass is heavier than about 200 GeV.
The lower limit of the stop in case of $m_{\tilde t_L}>m_{\tilde t_R}$ is larger than in
the case of  $m_{\tilde t_L}<m_{\tilde t_R}$. The reason is that
for $m_{\tilde{t}_L} > m_{\tilde{t}_R}$ the Higgs mass will receive a negative correction
proportional to $\cos2\beta\log(m_{\tilde{t}_L}/m_{\tilde{t}_R})$ \cite{higgs-mass}
and, in order to enhance the Higgs mass to 125 GeV, the splitting between
$m_{\tilde{t}_L}$ and $m_{\tilde{t}_R}$ is preferred to be small and
a heavier stop mass is favored to balance the negative correction.
Secondly, the masses of $\tilde{\chi}^{+}_{1}$, $\tilde{\chi}^{0}_{1}$ and $\tilde{\chi}^{0}_{2}$ are below 200 GeV
and are nearly degenerate, which, due to the soft decay products, will weaken the constraints from direct
electroweak gauginos searches at the LHC. However, this spectrum may strengthen some limits from direct
stop/sbottom searches at the LHC since the final states $b \tilde{\chi}^{+}_{1}$ ($t \tilde{\chi}^{-}_{1}$)
have the same signatures as $b \tilde{\chi}^{0}_{1,2}$ ($t \tilde{\chi}^{0}_{1,2}$).

\section{Constraints from the LHC search of stop/sbottom}
Since the decay modes of stop and sbottom will affect the implication of direct searches,
we first discuss their decay branching ratios.
The relevant interactions between stop and neutralinos or charginos are given by \cite{mssm-feynrules}
\begin{eqnarray}
    {\cal L}_{t\bar{t}\tilde{\chi}^0} &=& \tilde{t}_1
    \bar{t} ( f^{N}_L P_L + f^{N}_R P_R ) \tilde{\chi}^0_i + h.c.~, \\
    {\cal L}_{t\bar{b}\tilde{\chi}^\pm} &=& \tilde{t}_1
    \bar{b} ( f^{C}_L P_L + f^{C}_R P_R ) \tilde{\chi}^+_i +h.c.~,
    \label{vertex}
\end{eqnarray}
where $P_{L/R}=(1\mp\gamma_5)/2$ and
\begin{eqnarray}
    f^N_L &=&
    -\left[ \frac{1}{\sqrt{2}} g_2 [ N_{\chi^0} ]_{i \tilde{W}}
        + \frac{1}{3\sqrt{2}} g_1 [ N_{\chi^0} ]_{i \tilde{B}}
    \right] \cos\theta_{\tilde{t}}
    - y_t [ N_{\chi^0} ]_{i \tilde{H}_2} \sin\theta_{\tilde{t}},
\label{eq7} \\
    f^N_R &=& \frac{2\sqrt{2}}{3} g_1
    [ N_{\chi^0} ]^*_{i \tilde{B}} \sin\theta_{\tilde{t}}
    - y_t [ N_{\chi^0} ]^*_{i \tilde{H}_2} \cos\theta_{\tilde{t}},\\
        f^C_L &=& y_b [ U_{\chi^+} ]^{*}_{i \tilde{H}^-_1} \cos\theta_{\tilde{t}},\\
    f^C_R &=& - g_2[ V_{\chi^+} ]_{i \tilde{W}^{-}} \cos\theta_{\tilde{t}}
    + y_t[ V_{\chi^+} ]_{i \tilde{H}^+_2}\sin\theta_{\tilde{t}}.
\label{eq10}
\end{eqnarray}
with $y_t=\sqrt{2}m_t/(v\sin\beta)$ and $y_b=\sqrt{2}m_b/(v\cos\beta)$ being
the Yukawa couplings of top and bottom quarks, and $\theta_{\tilde{t}}$ being
the mixing angle between left- and right-handed stops
($-\pi/2 \leq \theta_{\tilde{t}} \leq \pi/2$).
The neutralino and chargino mixing matrices $N_{ij}$, $U_{ij}$ and $V_{ij}$
are defined in \cite{mssm-feynrules}.
The interactions of the sbottom with neutralino or chargino can be
obtained from above expressions
 by replacing $\theta_{\tilde{t}}$ with $\theta_{\tilde{b}}$
and interchanging $y_t$ and $y_b$.
\begin{figure}[h!]
\centering
\includegraphics[width=5in,height=2.5in]{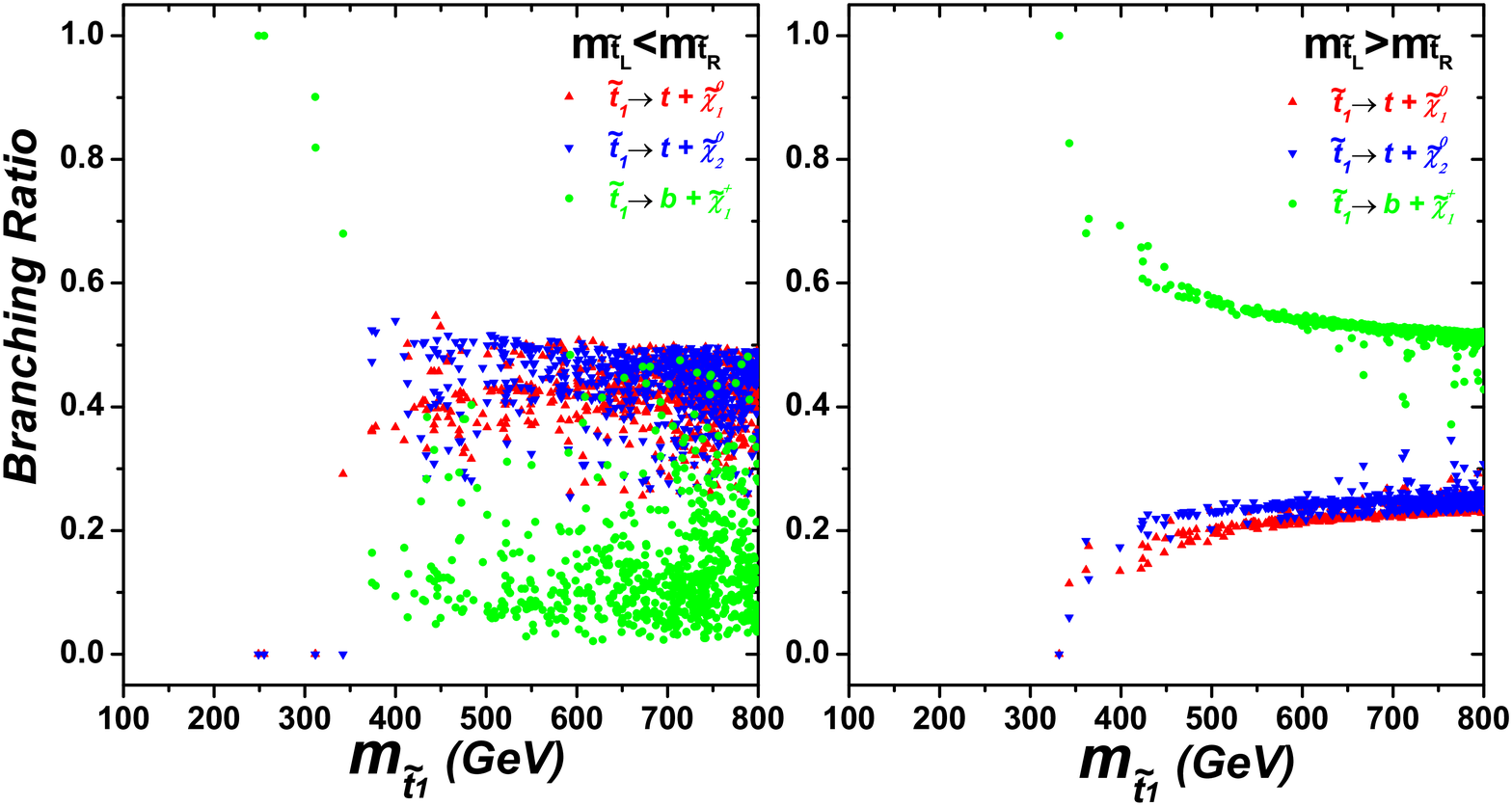}
\includegraphics[width=5in,height=2.5in]{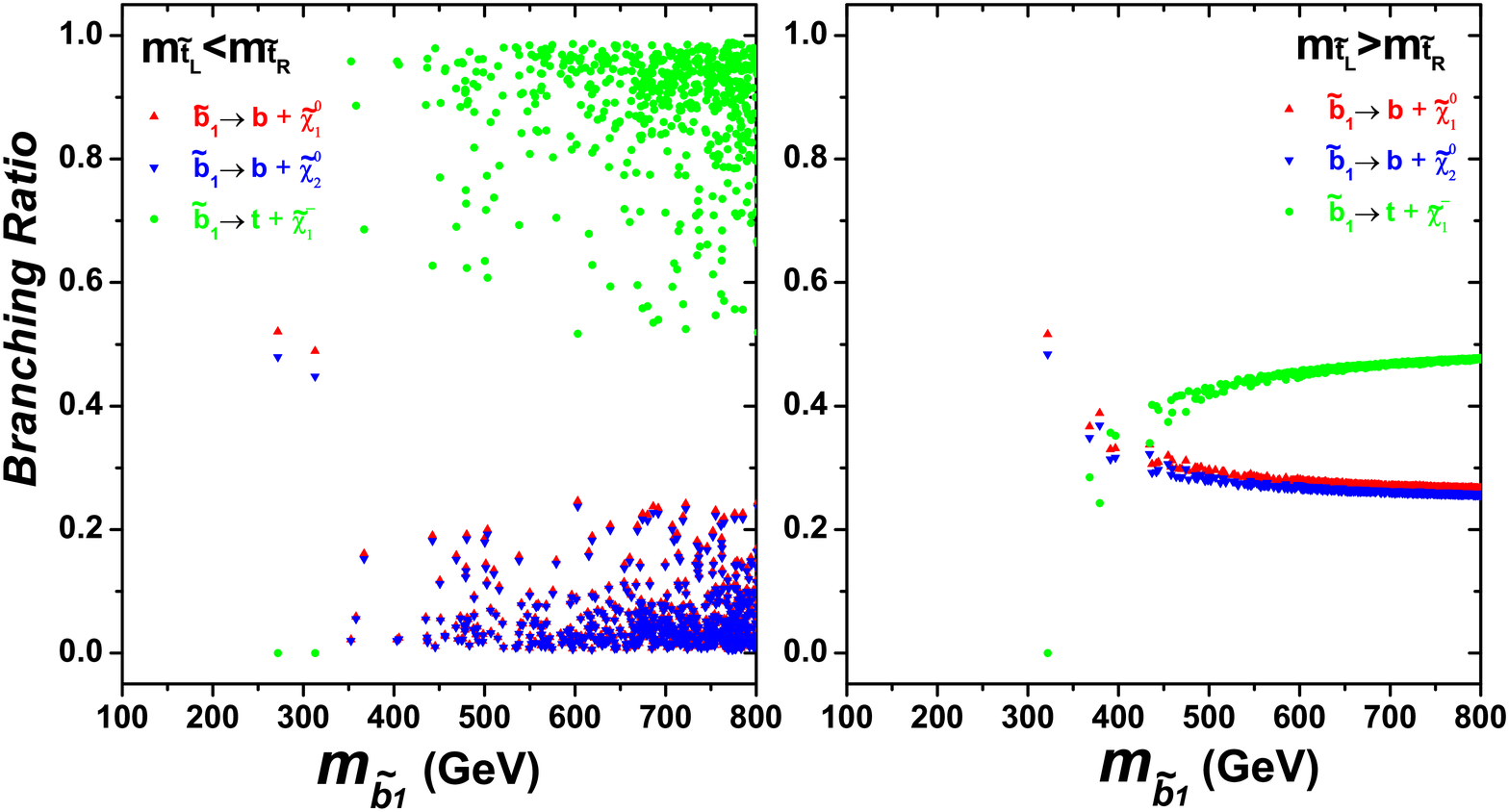}
\vspace*{-0.5cm}
\caption{The scatter plots of the parameter space allowed by constraints (1-4) in natural SUSY,
showing the decay branching ratios of stop and sbottom.}
\label{decays}
\end{figure}

In Fig.~\ref{decays} we display the numerical results of the decay branching ratios of stop and sbottom.
From the upper panel we can see that a light stop ($\tilde{t}_{1}<370$ GeV) will be dominated by decaying
to $b \tilde{\chi}^{+}_{1}$. When the stop becomes heavier, the $\tilde{t}_{1} \to t \tilde{\chi}^{0}_{1,2}$ channels
are open and the decay $\tilde{t}_1 \to b \tilde{\chi}^{+}_{1}$ is suppressed.
For $m_{\tilde{t}_L}< m_{\tilde{t}_R}$, ${\rm Br}(\tilde{t}_1 \to b \tilde{\chi}^{+}_{1})$
is less than 50\%, and ${\rm Br}(\tilde{t}_{1} \to t \tilde{\chi}^{0}_{1})$
and ${\rm Br}(\tilde{t}_{1} \to t \tilde{\chi}^{0}_{1,2})$
are approximately equal which vary between 25\% and 50\%.
While for $m_{\tilde{t}_L}> m_{\tilde{t}_R}$,
${\rm Br}(\tilde{t}_1 \to b \tilde{\chi}^{+}_{1})$ is around 50\%
and ${\rm Br}(\tilde{t}_{1} \to t \tilde{\chi}^{0}_{1,2})$ can only reach 25\%.
These features can be understood from Eqs.~(\ref{eq7}--\ref{eq10})
in the chiral limit.
From those expressions
we can see that $\tilde{t}_1$ becomes purely left-handed for $\theta_{\tilde{t}}=0$
and right-handed for $\theta_{\tilde{t}}=\pm \pi/2$.
Since $\tilde{\chi}^0_{1,2}$ are  higgsino-like, the couplings $C_{t\tilde{t}_{1} \tilde{\chi}^0_{1,2}}$
are proportional to $y_t$ and nearly equal for both  $m_{\tilde{t}_L}< m_{\tilde{t}_R}$ and
 $m_{\tilde{t}_L}> m_{\tilde{t}_R}$.
However, the coupling $C_{b\tilde{t}_1 \chi^+_1}$
is dominated by  the bottom Yukawa coupling $y_b$ for  $m_{\tilde{t}_L}< m_{\tilde{t}_R}$
and by the top Yukawa coupling $y_t$ for  $m_{\tilde{t}_L}> m_{\tilde{t}_R}$.
So, we can infer that when $\tilde{t}_{1} \to t \tilde{\chi}^{0}_{1,2}$ is kinematically allowed,
$Br(\tilde{t}_1 \to b \tilde{\chi}^{+}_{1})$ will be generally below $Br(\tilde{t}_{1} \to t \tilde{\chi}^{0}_{1,2})$
for $m_{\tilde{t}_L}< m_{\tilde{t}_R}$ and twice as large as $Br(\tilde{t}_{1} \to t \tilde{\chi}^{0}_{1,2})$
for $m_{\tilde{t}_L}> m_{\tilde{t}_R}$.
For the sbottom decays on the lower panel, they can be understood by similar analysis as above.

\begin{table}[t]
\caption{The signals of the ATLAS stop/sbottom direct searches \cite{exp-st,exp-sb}.\label{tab1}}
\begin{tabular}{|c|c|}
\hline~~ Signals of ATLAS stop/sbottom direct searches~~&
~~Source of each signal in natural SUSY~~\\
\hline
$\ell+jets+\slashed E_{T}$~ &
$
\begin{array}{ll} pp\to \tilde{t}_1 \tilde{t}_1~ (\tilde{t}_1 \to t \tilde{\chi}^{0}_{1,2}) \\
                    pp\to \tilde{b}_1 \tilde{b}_1~ (\tilde{b}_1 \to t \tilde{\chi}^{-}_{1})
   \end{array}
$
 \\ \hline
$t\bar{t}$(hadronic)+$\slashed E_{T}$ &
$
\begin{array}{ll}
   pp\to \tilde{t}_1 \tilde{t}_1~ (\tilde{t}_1 \to t \tilde{\chi}^{0}_{1,2})\\
   pp\to \tilde{b}_1 \tilde{b}_1~ (\tilde{b}_1 \to t \tilde{\chi}^{-}_{1})
   \end{array}
$
 \\ \hline
$2b+\slashed E_{T}$ &
$
\begin{array}{ll}
  pp\to \tilde{t}_1 \tilde{t}_1~ (\tilde{t}_1 \to b \tilde{\chi}^{+}_{1})\\
  pp\to \tilde{b}_1 \tilde{b}_1~ (\tilde{b}_1 \to b \tilde{\chi}^{0}_{1,2}) \\
   \end{array}
$ \\
\hline
\end{tabular}
\end{table}
\begin{figure}[h!]
\includegraphics[width=6in,height=2.5in]{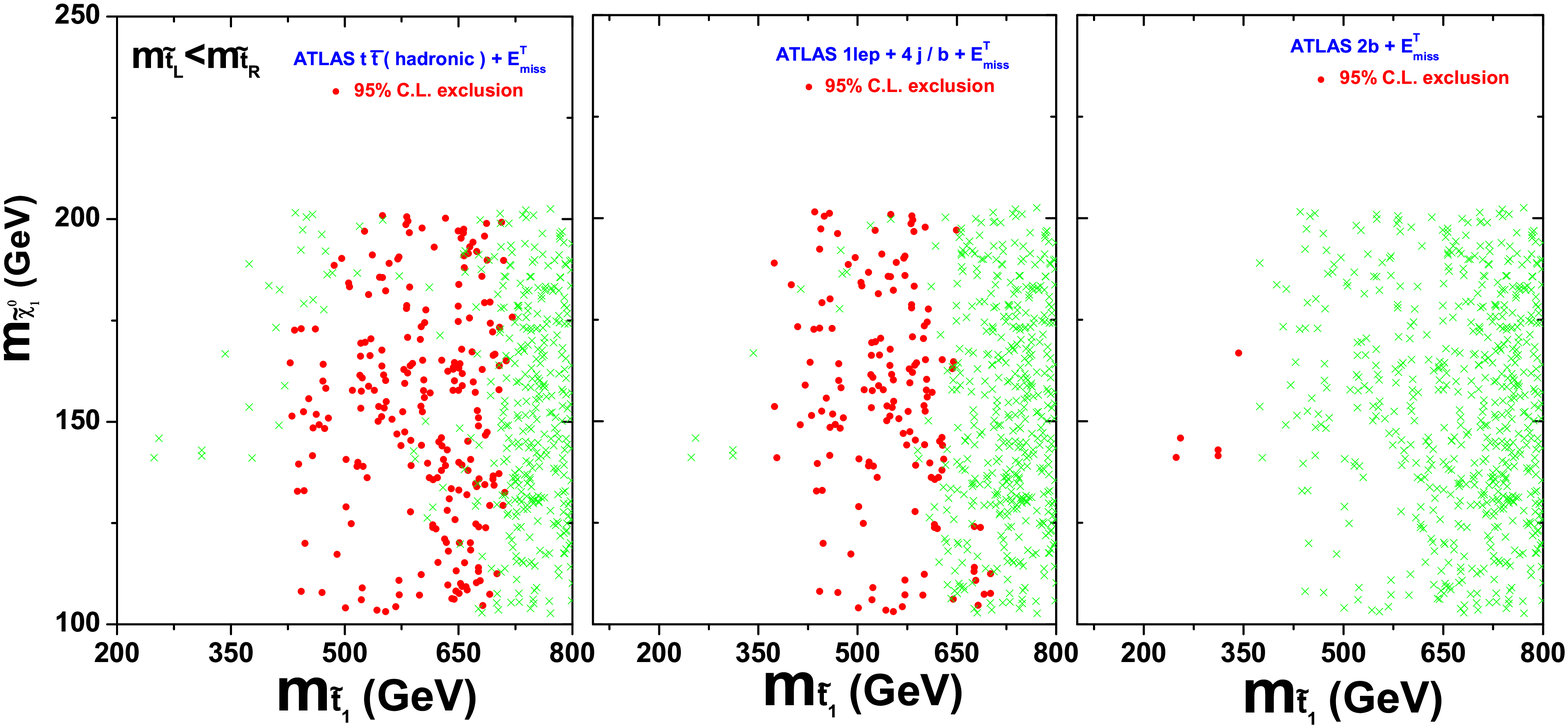}
\includegraphics[width=6in,height=2.5in]{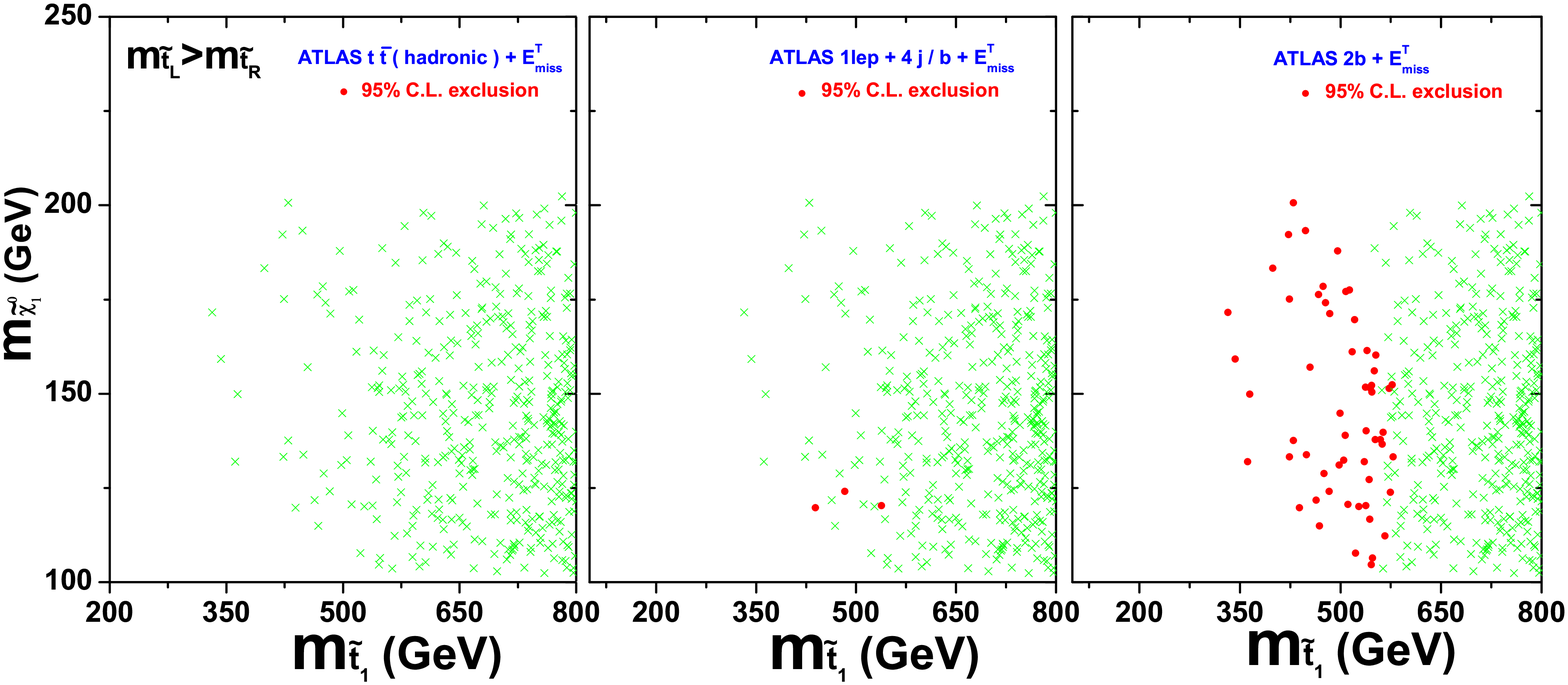}
\vspace*{-0.5cm}
\caption{Same as Fig.~\ref{decays}, but projected on the plane of stop mass versus neutralino mass.
The samples denoted by bullets (red) are excluded by the ATLAS
direct stop/sbottom search data at 95\% CL.}
\label{fig3}
\end{figure}

In the following, we use the ATLAS direct stop/sbottom search data \cite{exp-st,exp-sb} to
constrain the parameter space of natural SUSY.
The signals of the direct searches and the source of each signal in natural SUSY
are shown in Table I.
In our calculation we simulate the signals by \textsf{MadGraph5} \cite{mad5} and
carry out the parton shower and fast detector simulation with \textsf{PYTHIA} \cite{pythia}
and \textsf{Delphes} \cite{delphes}. We use the anti-$k_t$ algorithm \cite{anti-kt} to cluster jets
and the MLM scheme \cite{mlm} to match our matrix element with parton shower.
For the cross section of $\tilde{t}\tilde{t}^*$/$\tilde{b}\tilde{b}^*$ we use
the NLO results calculated by the package \textsf{Prospino2.1} \cite{prospino}.
Before presenting the exclusion limits,
in Appendix A we validate our simulations by adopting the same cuts
flow of ATLAS and comparing our results to the ATLAS results.
We can see that our results are consistent with ATLAS in the reasonable error ranges.
Then, we perform the simulation for each survived sample in Fig.~\ref{decays}
and use ${\rm CL}_s$ method to obtain the 95\% exclusion limit.
For each channel $i$ after cuts flow, we can define ${\rm CL}^i_s$ as
\begin{eqnarray}
{\rm CL}^i_s=\frac{\mathop{\rm Pois}(n_i|b_i+s_{i})}{\mathop{\rm Pois}(n_i|b_i)}.
\end{eqnarray}
Here $s_{i}$ is the signal events from our Monte Carlo simulation, and
$n_i$ and $b_i$ are  respectively the number of the observed and expected SM background events,
which can be taken from the experimental papers.
To account for the systematic uncertainty effect, we allow the number of background events
in each channel to fluctuate: $b_i \rightarrow b_i (1 + \delta b_i)$.
Following the analysis in \cite{conway}, we convolve the Poisson distribution with a Gaussian weighting
factor and then integrate over $b_i$:
\begin{eqnarray}
\mathop{\rm Pois}(n_i|b_i+s_{i})\rightarrow \int{d(\delta b_i)}\mathop{\rm Gaus}(\delta b_i,f_i^b)\mathop{\rm Pois}(n_i|b_i (1 + \delta b_i)+s_{i})
\end{eqnarray}
where $f^b_i$ is the width of the Gaussian distribution and is treated as the relative systematic uncertainty.
Besides, we truncate the Gaussian integration in the case of the negative events number of background.

In Fig.~\ref{fig3} we show the constraints of the ATLAS stop/sbottom search data on the parameter space
of natural SUSY. Generally, we can see that the stop mass
can be more stringently constrained in case of $m_{\tilde{t}_{L}} <  m_{\tilde{t}_{R}}$
than in case of $m_{\tilde{t}_{L}} >  m_{\tilde{t}_{R}}$,  which can be understood from
Fig.~\ref{decays} and Table \ref{tab1}.

In the former case, since $\tilde{t} \to t \tilde{\chi}^{0}_{1,2}$ and $\tilde{b} \to t \tilde{\chi}^{-}_{1}$
are the dominant decay modes for moderate stop/sbottom mass ($m_{\tilde{t}_1}>350$ GeV),
by using the ATLAS data of hadronic $t\bar{t}+\slashed E_{T}$ and $\ell+{\rm jets}+\slashed E_{T}$,
we can almost exclude $350 < m_{\tilde{t}_{1}} < 590$ GeV and $350 < m_{\tilde{t}_{1}} < 605$ GeV, respectively.
But we should mention that since  $\tilde{t} \to t \tilde{\chi}^{0}_{1,2}$ and $\tilde{b} \to t \tilde{\chi}^{-}_{1}$
have the same signal, some samples with a heavier stop can also be excluded by these two channels.
But in the latter case, the stop mass is hardly bounded by the stop searches due to the suppression
of $\tilde{t} \to t \tilde{\chi}^{0}_{1,2}$ and $\tilde{b} \to t \tilde{\chi}^{-}_{1}$.
Besides, we note that the sbottom search data of $2b+\slashed E_{T}$ can give a complementary constraint
on the stop mass, which excludes a light stop between 235--350 GeV  in the former case
and the exclusion limit can be up to 545 GeV in the latter case.

\begin{figure}[h!]
\centering
\includegraphics[width=6in,height=2.8in]{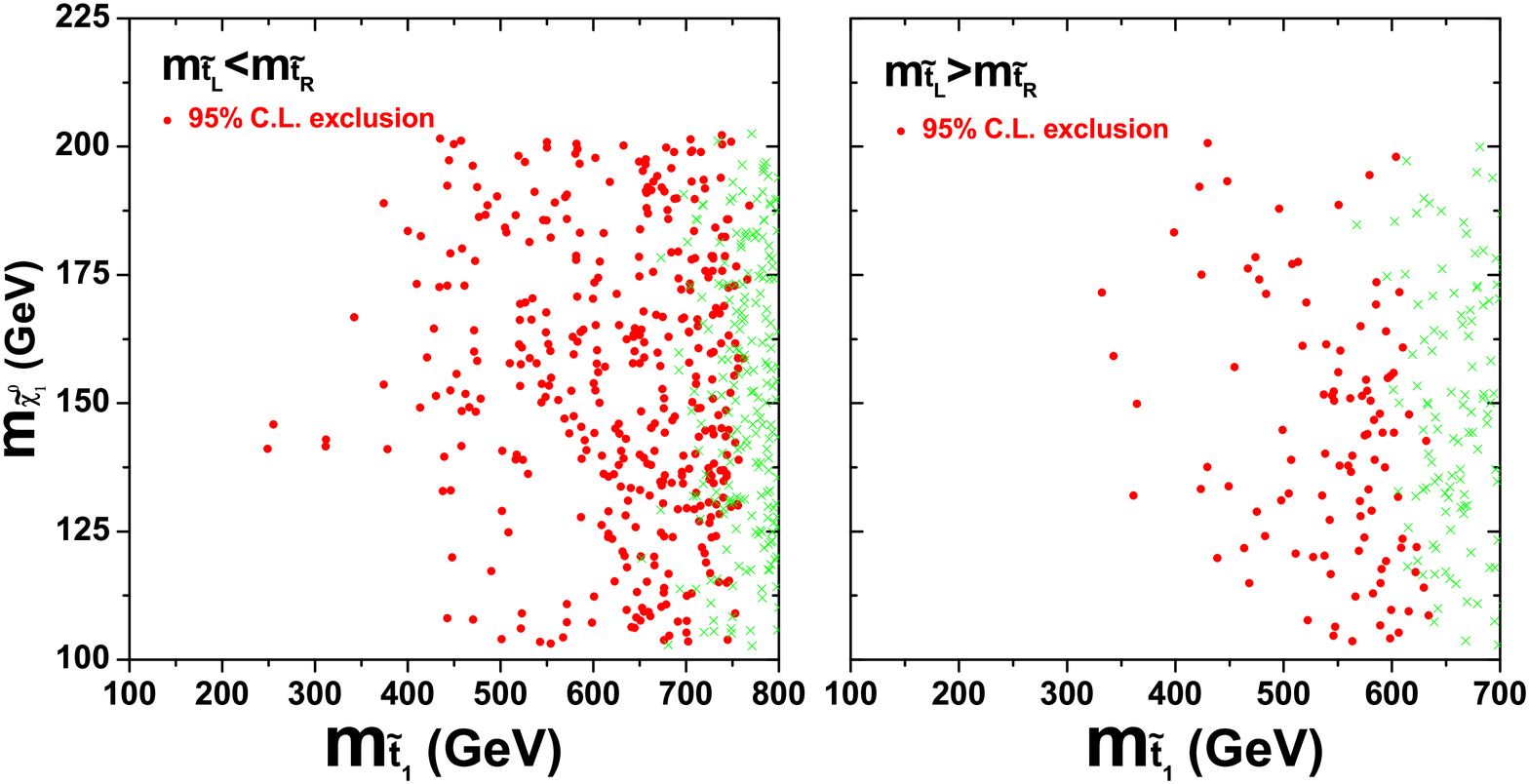}
\vspace*{-0.5cm}
\caption{Same as Fig.~\ref{fig3}, but showing the combined exclusion limits.}
\label{fig4}
\end{figure}

In Fig.~\ref{fig4} we combine the above exclusion limits from the three channels with $CL_s$ method.
It can be seen that the data can exclude at 95\% CL a stop between 200--670 GeV for
$m_{\tilde{t}_{L}} <  m_{\tilde{t}_{R}}$ and 250--600 GeV for $m_{\tilde{t}_{L}} >  m_{\tilde{t}_{R}}$.
These combined limits are much stronger than the limits from single channel.

Note that since our analysis aim at the natural SUSY where the value of $\mu$ is preferred to be small (usually below 200 GeV),
our results are relevant for the low $\mu$. If $\mu$ becomes large(the lightest neutralino becomes heavy)
but the stop mass is fixed, the small mass difference between them will make the visible decay products
(jets or leptons) of the stop be soft. Thus the signal
detection efficiency becomes lower so that our direct search limits will become weak.
We checked and found that when $m_{\tilde{t}_1}=500$ GeV, $\mu > 300$ GeV can escape the constraints
considered in this work.Besides, we noticed that if $\mu$ gets heavier, the stop should also become
heavy to keep the neutralino as LSP, which will suppress the production rate of stop pair and escape our constraints.

\begin{figure}[h!]
\centering
\includegraphics[width=4in]{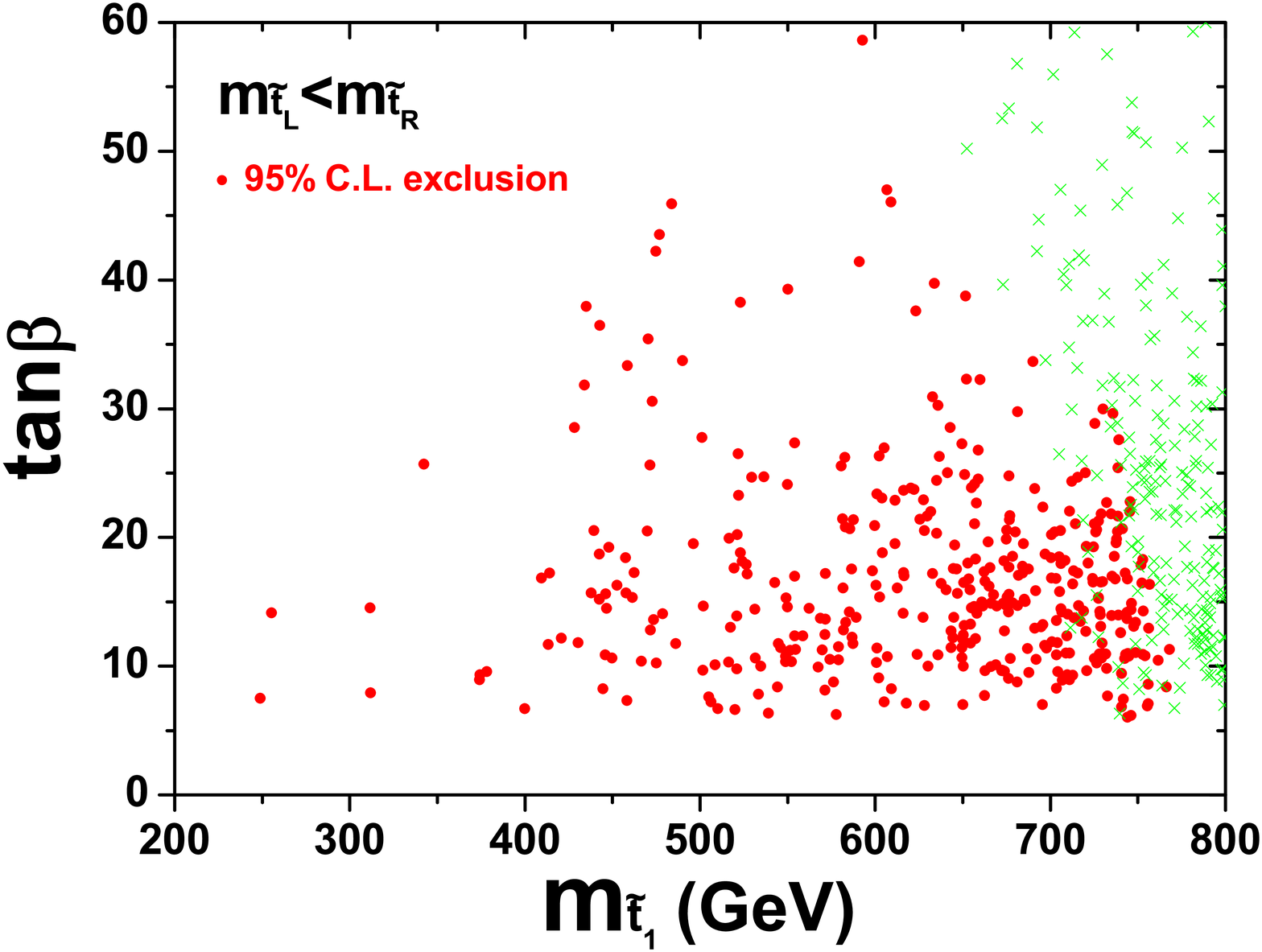}
\vspace*{-0.5cm}
\caption{Same as Fig.~\ref{fig4}, but showing the dependence on
$\tan\beta$.}
\label{fig5}
\end{figure}

The effects of $\tan\beta$ on our results are shown in Fig.\ref{fig5}
in the case of $m_{\tilde t_L}<m_{\tilde t_R}$.
From this figure we see that the limits get weak for a
large $\tan\beta$. The reason is that in the case of $m_{\tilde t_L}<m_{\tilde t_R}$
the decay width of $\tilde t_1\to \tilde\chi^+_1 b$
($\tilde t_1\to t \tilde\chi^0_1$) is determined by
$y_b \sim 1/\cos\beta$ ($y_t \sim 1/\sin\beta$), and thus when $\tan\beta$ gets large,
the branching ratio of $\tilde t_1\to t \tilde\chi^0_1$ will be suppressed. Then the
limits on the stop mass, which dominantly come from the decay
$\tilde t_1\to t \tilde\chi^0_1$, will become weak.

In the case of $m_{\tilde t_L}>m_{\tilde t_R}$, our results are not sensitive
to the value of $\tan\beta$
because in this case the decay widths of $\tilde t_1\to \tilde\chi^+_1 b$
and $\tilde t_1\to t \tilde\chi^0_1$ are both determined by $y_t$ and their
branching ratios are not sensitive to $\tan\beta$.

\begin{figure}[h!]
\centering
\includegraphics[width=6in,height=2.8in]{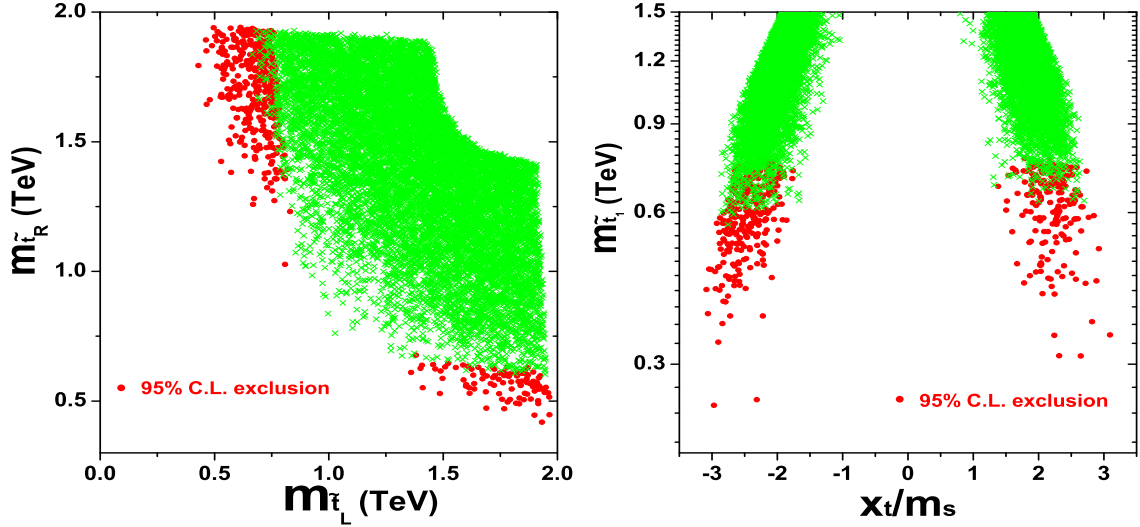}
\vspace*{-0.5cm}
\caption{Same as Fig.~\ref{fig3}, but projected on different planes .}
\label{fig6}
\end{figure}

In Fig.~\ref{fig6} we display the impact of the direct stop/sbottom searches on the stop sector. From the left panel we see that the two regions can be
excluded by direct searches at 95\% CL.
The right panel shows that the range of $2<X_t/M_s<3$ can be excluded
for $m_{\tilde{t}_1}<600$ GeV at 95\% CL.

Since the decay branching ratios of stop will be affected by the parameter $A_t$ through the mixing angle $\theta_{\tilde{t}}$,
we let $A_t$ vary in our calculations to examine the its effect on the exclusion limit. However, when the stop mass $m_{\tilde{t}_1}$ is low
(so that the LHC data can set limits), in order to get a 125 GeV SM-like Higgs we need a large $m_{\tilde{t}_2}$ in addition to a large $A_t$.
Then from the formula $\sin(2\theta_{\tilde{t}})=2m_tX_t/(m^2_{\tilde{t}_1}-m^2_{\tilde{t}_2})$ where $X_t=A_t-\mu/\tan\beta$,
we see that the value of $\sin(2\theta_{\tilde{t}})$ is small and thus $\theta_{\tilde{t}}$ will be close to 0 or $\pm\pi/2$,
which means that the stop is highly left-handed or right-handed. So the variation of $A_t$ can affect the stop mixing, but
its effect is quite limited due to our requirement of 125 GeV Higgs.
We checked numerically that the constraints are not sensitive to the value of $A_t$ in our case and
mainly depend on the light stop scale and $\tan\beta$.

\section{Conclusion}
We studied the constraints of the LHC search results on the stop mass
in natural SUSY. Considering the constraints from the Higgs mass, B-physics and
electroweak precision measurements, we scanned the parameter space of natural SUSY in the framework of MSSM,
and then in the allowed parameter space we performed a Monte Carlo simulation for stop pair production
followed by $\tilde{t}_{1} \to t \tilde{\chi}_{1}^{0}$ or $\tilde{t}_{1} \to b \tilde{\chi}_{1}^{+}$
and sbottom pair production followed by $\tilde{b}_{1} \to b \tilde{\chi}_{1}^{0}$ or
$\tilde{b}_{1} \to t \tilde{\chi}_{1}^{-}$.
Using the combined results of ATLAS with 20.1 fb$^{-1}$ from
the search of $\ell+{\rm jets}+\slashed E_{T}$, hadronic $t\bar{t}+\slashed E_{T}$ and $2b+\slashed E_{T}$,
we found that a stop lighter than 600 GeV can be excluded at 95\% CL.

\section*{Acknowledgement}
We appreciate the helpful discussions with Archil Kobakhidze, Junjie Cao and Jie Ren.
This work was supported in part by
the ARC Center of Excellence for Particle Physics at the Tera-scale,
by the National Natural Science Foundation of China (NNSFC)
under grant No. 11275245, 10821504, 11135003, 11305049 and 11275057,
by the Startup Foundation for Doctors of Henan Normal University under contract No.11112, and by the Grant-in-Aid for Scientific Research (No.~24540246)
from Ministry of Education, Culture, Sports, Science and Technology (MEXT)
of Japan.

\appendix

\section{Comparison of our simulation with the ATLAS results}
In the following we present the comparison of our simulation with the ATLAS results.
In Tables \ref{tab2}, \ref{tab3} and \ref{tab4} we present the results for
$2b+\slashed E_{T}$,  $\ell+{\rm jets}+\slashed E_{T}$ and hadronic $t\bar{t}+\slashed E_{T}$,
respectively.

\begin{table}[h!] \caption{Comparison of ATLAS and our simulation for $2b+\slashed E_{T}$ when ($\tilde{b}_1,\tilde{\chi}^0_1$)=(600,1)GeV \label{tab2}}
\begin{tabular}{|c|c|c|c|c|c|}
\hline cuts ~&~ATLAS~&~our simulation\\
\hline ~ ~~$\slashed E_{T}$ $>$ 150 (GeV) ~~ & ~~ 270 ~~   &~~ 239.27  \\
\hline ~ ~~$P_T(j_1),P_T(j_2),j_3$(veto) ~~ & ~~ 110 ~~ &~~ 91.85  \\
\hline ~ ~~$\Delta\phi_{\rm min}(3)$ ~~ & ~~ 93 ~~ &~~ 73  \\
\hline ~ ~~$\slashed E_{T}^{\rm miss}/m_{\rm eff}(j_1,j_2,j_3)$ and b-tag ~~ & ~~ 24 ~~ &~~ 19.46  \\
\hline ~ ~~$m_{CT} > 100 $ (GeV)  ~~ & ~~ 22 ~~ &~~ 17.27 \\
\hline
\end{tabular}
\end{table}

\begin{table}[h!]
\caption{Comparison of ATLAS and our simulation for $\ell+{\rm jets}+\slashed E_{T}$ when ($\tilde{t}_1,\tilde{\chi}^0_1$)=(650,1)GeV \label{tab3}}
\begin{tabular}{|c|c|c|c|c|c|}
\hline cuts ~&~ATLAS~&~our simulation\\
\hline ~ ~~No selection ~~ & ~~ 500004.0 ~~   &~~ 500004.0  \\
\hline ~ ~~1 $\mu$ and 4 jets (80,60,40,25)  ~~ & ~~  3704 ~~ &~~  3468  \\
\hline ~ ~~ $\geq$ 1 b-tag in 4 leading jets ~~ & ~~ 3145.2 ~~ &~~ 3089  \\
\hline ~ ~~$\slashed E_{T}^{\rm miss}> 100$ (GeV)   ~~ & ~~ 2930  ~~ &~~ 2838   \\
\hline ~ ~~$\slashed E_{T}^{\rm miss}/\sqrt{H_T} > 5$ ~~ & ~~ 2844.2 ~~ &~~ 2757 \\
\hline ~ ~~$\Delta\phi({\rm jet}, \slashed E_{T}^{\rm miss}) > 0.8$  ~~ & ~~ 2649 ~~ &~~ 2488 \\
\hline ~ ~~$\slashed E_{T}^{\rm miss}> 200$ (GeV)   ~~ & ~~ 2210.6  ~~ &~~ 2016   \\
\hline ~ ~~$\slashed E_{T}^{\rm miss}/\sqrt{H_T} > 13$ ~~ & ~~ 1619.1 ~~ &~~ 1382 \\
\hline ~ ~~$m_{T} > 140 $ (GeV)  ~~ & ~~ 1474.9 ~~ &~~ 1203 \\
\hline
\end{tabular}
\end{table}

\begin{table}[h!] \caption{Comparison of ATLAS and our simulation for hadronic $t\bar{t}+\slashed E_{T}$ when ($\tilde{t}_1,\tilde{\chi}^0_1$)=(600,0)GeV \label{tab4}}
\begin{tabular}{|c|c|c|c|c|c|}
\hline cuts ~&~ATLAS~&~our simulation\\
\hline ~ ~~No selection ~~ & ~~ 507.5 ~~   &~~ 507.5  \\
\hline
$
\begin{array}{ll}
       ~~~e/\mu ~~\rm{veto}     \\
       \slashed E_{T}^{\rm miss} >130 \rm{(GeV)}
\end{array}
$
~~ & ~~ 270.1 ~~&~~ 303.8  \\
\hline ~ ~~Jet multiplicity and $p_T$ ~~ & ~~ 92.2 ~~ &~~ 105.1  \\
\hline ~ ~~$\slashed E_{T}^{\rm track}> 30$ (GeV)   ~~ & ~~   ~~ &~~    \\
       ~ ~~$\Delta\phi(jet,\slashed E_{T}^{\rm miss}) > \pi/5$    ~~ & ~~ 72 ~~ &~~ 82.5  \\
       ~ ~~ $\Delta\phi(\slashed E_{T}^{\rm miss},\slashed E_{T}^{\rm miss,track}) < \pi/3$ ~~ & ~~   ~~ &~~  \\
\hline ~ ~~tau veto and b-tag  ~~ & ~~ 31.5 ~~ &~~ 33.06 \\
\hline ~ ~~$m_T(\mbox{$b$-jet}, \slashed E_{T}^{\rm miss}) > 175$ (GeV)  ~~ & ~~ 23.6 ~~ &~~ 19.3 \\
\hline ~ ~~80 (GeV) $< m_{jjj}^{0,1} < 270$ (GeV) ~~ & ~~ 11.9 ~~ &~~ 10.14 \\
\hline ~ ~~$E_{T} > 150 $ (GeV)  ~~ & ~~ 11.8 ~~ &~~ 9.89 \\
\hline
\end{tabular}
\end{table}
\newpage
\hbox to \hsize{\hss}

\end{document}